\documentclass[aps,
	       prd,
	       nofootinbib,
	       preprintnumbers,
	       nobibnotes,
	       twocolumn]{revtex4-1} %twocolumn

\usepackage{amssymb}
\usepackage{amsmath}
\usepackage{graphicx,subfigure}
\usepackage{longtable}
\usepackage{verbatim}
\usepackage{amsfonts}
\usepackage{hyperref}
\usepackage{color}

\newcommand{\be}{\begin{equation}}
\newcommand{\ee}{\end{equation}}
\newcommand{\bea}{\begin{eqnarray}}
\newcommand{\eea}{\end{eqnarray}}
\newcommand{\beq}{\begin{equation}}
\newcommand{\eeq}{\end{equation}}
\def\beqa{\begin{eqnarray}}
  \def\eeqa{\end{eqnarray}}
\newcommand{\bv}{\left(\begin{array}{c}}
\newcommand{\ev}{\end{array}\right)}

\def\lsim{\mathrel{\rlap{\lower4pt\hbox{\hskip1pt$\sim$}}
    \raise1pt\hbox{$<$}}}	  %less than or approx. symbol
\def\gsim{\mathrel{\rlap{\lower4pt\hbox{\hskip1pt$\sim$}}
    \raise1pt\hbox{$>$}}}	  %greater than or approx. symbol

\newcommand{\nn}{\nonumber}

\begin{document}

%\preprint{ }

\title{A fresh look at the determination of $|V_{cb}|$ from
 $B\to D^{*}l\nu$}
\author{Dante Bigi} 
\email{dante.bigi@to.infn.it}
\author{Paolo Gambino}
\email{gambino@to.infn.it}
\author{Stefan Schacht}
\email{schacht@to.infn.it}
\affiliation{
Dipartimento di Fisica, Universit\`a di Torino \& INFN, Sezione di Torino, I-10125 Torino, Italy}

\vspace*{1cm}

\begin{abstract}
We use recent Belle results on $\bar{B}^0\rightarrow D^{*+}l^-\bar{\nu}_l$ decays
to extract the CKM element 
$|V_{cb}|$ with two different but well-founded parameterizations of the  form factors. 
We show that the CLN and BGL parameterizations lead to quite different 
results for $|V_{cb}|$ and provide a simple explanation of this unexpected behaviour. A long lasting discrepancy between the inclusive and exclusive determinations of $|V_{cb}|$ may have  to be thoroughly reconsidered.

\end{abstract}

\maketitle

\section{Introduction}

Semileptonic $B$ decays offer the most direct way to determine the  element $|V_{cb}|$ of
the Cabibbo-Kobayashi Maskawa (CKM) quark mixing matrix. This particular element
plays a central role in the analyses of the CKM matrix unitarity and in the SM prediction of Flavour Changing Neutral Current transitions. For a long time the two available methods to extract $|V_{cb}|$ from experimental data, based on exclusive (single hadronic channel) and inclusive (sum of all hadronic channels) reconstruction of the semileptonic $B$ decays, have been in conflict. The two methods are based on very different theoretical foundations and while a new physics interpretation seems currently disfavoured on general grounds \cite{Crivellin:2014zpa}, it is not excluded  \cite{Colangelo:2016ymy} and is particularly interesting in view of the anomalies in $B\to D^{(*)} \tau \nu$ \cite{Amhis:2016xyh}. 

At present, the two most precise determinations are
\be
|V_{cb}|= (38.71\pm 0.75)\ 10^{-3},\label{excl}
\ee
based on the HFAG global combination of $B\to D^*\ell\nu$ results \cite{Amhis:2016xyh} together with the
FNAL-MILC Collaboration calculation \cite{Bailey:2014tva} of the relevant form factor at zero-recoil,~\emph{i.e.},~when the $D^*$ is produced at rest in the $B$ rest frame,  and
\be
|V_{cb}|= (42.00\pm 0.65)\ 10^{-3},\label{incl}
\ee
obtained  in the Heavy Quark Expansion 
from a fit to the moments of various kinematic distributions in inclusive semileptonic decays \cite{Gambino:2016jkc}.
The difference between  (\ref{excl}) and (\ref{incl}) is 3.3$\sigma$, which becomes  3.1$\sigma$
once the QED corrections are treated in the same way in both cases. There are alternative calculations of the $B\to D^*$ zero-recoil form factor on the lattice \cite{Harrison:2016gup} or based on Heavy Quark Sum Rules \cite{Gambino:2010bp,Gambino:2012rd} but they have larger uncertainties.

In a recent paper \cite{Bigi:2016mdz} we have reviewed and slightly updated the 20 years-old formalism to
parameterize  the form factors in $B\to D\ell\nu$ in a way that satisfies important unitarity constraints. Using up-to-date lattice calculations of the form factors and the available experimental results, we have shown that the parameterization dependence is small and obtained $|V_{cb}|=40.49(97)\ 10^{-3}$, compatible with both (\ref{excl}) and (\ref{incl}) and only slightly less precise.

The purpose of this Letter is to perform a similar analysis for the $B\to D^*\ell\nu$ decay. We take advantage of the new Belle preliminary results \cite{Abdesselam:2017kjf} which, for the first time, include deconvoluted kinematic and angular distributions with complete statistical and systematic errors and correlations, without relying on a particular parameterization of the form factors.
We first review the formalism and the data and then describe our fits and discuss the results.

\section{Form factor parameterizations}
In the limit of massless leptons 
the fully differential decay rate is given by 
\bea
 \frac{d\Gamma(\bar{B}\rightarrow D^*l\bar{\nu}_l)}{dw\, d\cos\theta_v\, d\cos\theta_l\, d\chi} \!&\!=&\!
\frac{\eta_{\mathrm{EW}}^2 3 m_B m_{D^*}^2}{4 (4 \pi)^4} \sqrt{w^2-1} \times \nn\\
&& \!(1 - 2 w r + r^2) G_F^2 \vert V_{cb}\vert^2 \times\nn\\
&&\! \!\left\{
	(1 - c_l)^2 s_v^2 H_+^2 + (1 + c_l)^2 s_v^2 H_-^2 \right.\nn\\
&&\!\!	+ 4 s_l^2 c_v^2 H_0^2 - 2 s_l^2 s_v^2 \cos 2 \chi H_+ H_- \nn\\
&& \!\!- 4 s_l (1 - c_l) s_v c_v \cos\chi H_+ H_0 \label{eq:fully-diff-decay-rate}\\
&& \!\!\left.+ 4 s_l (1 + c_l) s_v c_v \cos\chi H_- H_0
\right\}, \nn
\eea
where $r=m_{D^*}/m_B$, $c_v\equiv \cos\theta_v$, $c_l\equiv \cos\theta_l$, and 
correspondingly for $\sin\theta_v$ and $\sin\theta_l$.  $\theta_{v,l}$ and $\chi$ are the three angles that characterise the semileptonic decay.
We also use the kinematic parameter 
\be
w = \frac{m_B^2+m_{D^*}^2-q^2}{2 m_B m_{D^*}}\,,
\ee
where $q^2$ is the invariant mass of the lepton pair.

The helicity amplitudes $H_{\pm,0}$ in Eq.~(\ref{eq:fully-diff-decay-rate})  are given in terms of three form factors, see {\it e.g.}\  
Eqs.~(3-5) of Ref.~\cite{Abdesselam:2017kjf}.
In the Caprini-Lellouch-Neubert (CLN) parameterization \cite{Caprini:1997mu} one employs the 
form factor $h_{A_1}(w)$ and the ratios $R_{1,2}(w)$. Traditionally, the experimental collaborations use 
\bea
h_{A_1}(w)&=&h_{A_1}(1)\left[ 1-8 \rho^2 z
+(53\rho^2-15)z^2\right.\nn\\&&\left.\qquad\qquad\qquad\qquad-(231\rho^2-91)z^3\right],\nn\\
R_1(w)&=&R_1(1) -0.12 (w-1)+0.05 (w-1)^2,\label{CLN}\\
R_2(w)&=&R_2(1)+0.11(w-1)-0.06(w-1)^2,\nn
\eea
where $z=(\sqrt{w+1}-\sqrt{2})/(\sqrt{w+1}+\sqrt{2})$ and there are four independent parameters in total. We will discuss the ingredients of this parameterization later on.
\begin{table}[t]
\begin{center}
\begin{tabular}{c|c|c}
\hline\hline
Type   & Mass (GeV)  & 	References \\\hline
$1^-$  & $6.329$     & \cite{Dowdall:2012ab} 		\\
$1^-$  & $6.920$     & \cite{Dowdall:2012ab} 		\\
$1^-$  & $7.020$     & \cite{Devlani:2014nda} 		\\
$1^-$  & $7.280$     & \cite{Eichten:1994gt} 		\\
$1^+$  &  $6.739$   & \cite{Dowdall:2012ab} 		\\
$1^+$  &  $6.750$   & \cite{Devlani:2014nda,Godfrey:2004ya} \\
$1^+$  &  $7.145$   & \cite{Devlani:2014nda,Godfrey:2004ya} \\
$1^+$  &  $7.150$   & \cite{Devlani:2014nda,Godfrey:2004ya} \\
\hline\hline
\end{tabular}
\caption{Relevant $B_c^{(*)}$ masses. The $1^-$ resonances are as in Ref.~\cite{Bigi:2016mdz}.
\label{tab:relevant-Bc-masses}} 
\end{center}
\end{table}
After integration  over the angular variables, the $w$ distribution is proportional to \cite{Caprini:1997mu}
%\begin{align}
%\frac{d\Gamma}{dw} &= \vert V_{cb}\vert^2\frac{\eta^2_{\mathrm{EW}}G_F^2}{48\pi^3} m^3_{D^*} (m_B-m_{D^*})^2 \mathcal{G}(w) \mathcal{F}^2(w)\,,
%\end{align}
%with 
\begin{align}
 \mathcal{F}^2(w)  &= h^2_{A_1}(w) %\sqrt{w^2 - 1} (w+1)^2 
  \left(
	1 + 4 \frac{w}{w+1} \frac{1 - 2 w r + r^2}{(1 - r)^2}
	\right)^{-1}
 \times\nn\\
	&\left[
	2\frac{1 - 2 w r + r^2}{(1 - r)^2} \left(1 + R_1^2(w) \frac{w - 1}{w + 1}\right) +\right. \nn\\ 
	&\quad\left.\left(1 + ( 1 - R_2(w) ) \frac{w - 1}{1 - r}\right)^2
	\right]\,.\label{calF}
\end{align} 

An alternative parameterization is due to Boyd, Grinstein and Lebed (BGL) \cite{Boyd:1997kz}. In their notation the helicity amplitudes $H_i$ are given by 
\begin{align}
H_0(w) &= {\cal F}_1(w)/\sqrt{q^2},\nn\\
H_{\pm}(w) &= f(w)\mp m_B m_{D^*} \sqrt{w^2-1} \,g(w).\nn
\end{align}
\begin{table}[t]
\begin{center}
\begin{tabular}{c|c}
\hline\hline
Input  		  & 	Value \\\hline
$m_{B^0}$	  	  &	5.280 GeV\\
$m_{D^{*+}}$  	  &	2.010 GeV\\
$\eta_{\mathrm{EW}}$ &  $1.0066$ \\
$\tilde\chi^T_{1^-}(0)$ & 	$5.131\cdot 10^{-4}$ GeV$^{-2}$ \\
$\chi^T_{1^+}(0)$ & 	$3.894\cdot 10^{-4}$ GeV$^{-2}$ \\
\hline\hline
\end{tabular}
\caption{Further numerical inputs (uncertainties are small and can be neglected). The calculation of $\tilde\chi^T_{1^-}(0)$ and $\chi^T_{1^+}(0)$ follows Refs.~\cite{Grigo:2012ji,Bigi:2016mdz}.
\label{tab:further-input}} 
\end{center}
\end{table}
The relations between the relevant form factors in the CLN and BGL notation are 
\bea
&f = \sqrt{m_B m_D^*} (1 + w)\, h_{A_1}\,, \qquad     g = {h_V}/{\sqrt{m_B m_D^*}}\,, \nn\\
&\mathcal{F}_1 = (1 + w) ( m_B - m_{D^*} ) \sqrt{m_B m_{D^*} } A_5\,,\nn
\eea
and
\bea
R_1(w)&=&(w+1) \,m_B m_{D^*} \frac{g(w)}{f(w)},\nn\\
R_2(w)&=& \frac{w-r}{w-1}-\frac{{\cal F}_1(w)}{m_B (w-1) f(w)}.\nn
\eea

The three BGL form factors can be written as series in $z$,
\begin{align}
f (z)&= \frac{1}{P_{1^+}(z) \phi_{f}(z)} \sum_{n=0}^\infty a^{f}_n z^n \,, \nn \\
{\cal F}_1(z) &= \frac{1}{P_{1^+}(z) \phi_{{\cal F}_1}(z)} \sum_{n=0}^\infty a^{{\cal F}_1}_n z^n\,,\label{BGLexp}\\
g(z) &= \frac{1}{P_{1^-}(z) \phi_{g}(z)} \sum_{n=0}^\infty a^{g}_n z^n.\nn
\end{align}
In these equations the Blaschke factors $P_{1^\pm}$ are given by  
\be
\label{p+0}
P_{1^\pm}(z)=\prod_{P=1}^{n}\frac{z-z_{P}}{1-zz_{P}}, 
\ee
where  $z_{P}$ is defined as ($t_\pm=(m_B\pm m_{D^*})^2$)
\begin{equation*}
z_{P}=\frac{\sqrt{t_{+}-m_{P}^2}-\sqrt{t_{+}-t_{-}}}{\sqrt{t_{+}-m_{P}^2}+\sqrt{t_{+}-t_{-}}},
\end{equation*}
and the product  is extended  
to all the $B_c$ resonances below the $B$-$D^*$ threshold (7.29\,GeV) with the appropriate quantum numbers ($1^+$ for $f$ and ${\cal F}_1$, and $1^-$ for $g$). We use the $B_c$ resonances reported 
in Table~\ref{tab:relevant-Bc-masses}, but do not include the fourth  $1^-$ resonance, which is too uncertain and close to threshold.
\begin{table*}[t]
\begin{center} 
\subtable[\label{tab:fit-results-formfactors-BGL}]{
\begin{tabular}{c|c|c}
\hline
\hline
BGL Fit: 		 & Data + lattice	& Data + lattice + LCSR 	\\\hline 
%%%
$\chi^2/\mathrm{dof}$    & $27.9/32$ 		& $31.4/35$				\\\hline
%%%
$\vert V_{cb}\vert$      & $0.0417\left(^{+20}_{-21}\right)$ 	& $0.0404\left(^{+16}_{-17}\right)$ 		\\\hline
%%%
$a_0^f$	         	 & $0.01223(18)$			& $0.01224(18)$ 				\\ 
$a_1^f$	         	 & $-0.054\left(^{+58}_{-43}\right)$    & $-0.052\left(^{+27}_{-15}\right)$		\\
$a_2^f$	         	 & $ 0.2\left(^{+7}_{-12}\right)$       & $1.0\left(^{+0}_{-5}\right)$			\\\hline
%%%
$a_1^{\mathcal{F}_1}$	 & $-0.0100\left(^{+61}_{-56}\right)$ 	& $-0.0070\left(^{+54}_{-52}\right)$ 		\\
$a_2^{\mathcal{F}_1}$	 & $0.12\left(10\right)$    		& $ 0.089\left(^{+96}_{-100}\right)$ 		\\\hline
%%%
$a_0^{g}$	         & $0.012\left(^{+11}_{-8}\right)$   & $0.0289\left(^{+57}_{-37}\right)$ 		\\
$a_1^{g}$	         & $0.7\left(^{+3}_{-4}\right)$     	& $0.08\left(^{+8}_{-22}\right)$		\\
$a_2^{g}$	         & $0.8\left(^{+2}_{-17}\right)$	& $-1.0\left(^{+20}_{-0}\right)$		\\\hline\hline
\end{tabular}
}
\subtable[\label{tab:fit-results-formfactors-CLN} ]{ 
\begin{tabular}{c|c|c}
\hline
\hline
CLN Fit: 		 & Data + lattice 			& Data + lattice + LCSR 	\\\hline 
%%%
$\chi^2/\mathrm{dof}$    & $34.3/36$			& $34.8/39$		\\\hline
%%%
$\vert V_{cb}\vert$      & $0.0382\left(15\right)$		& $0.0382\left(14\right)$		\\\hline
%%%
$\rho_{D^*}^2$	         & $1.17\left(^{+15}_{-16}\right)$	& $ 1.16\left(14\right)$	\\ 
$R_1(1)$	         & $1.391\left(^{+92}_{-88}\right)$	& $1.372\left(36\right)$		\\
$R_2(1)$	         & $0.913\left(^{+73}_{-80}\right)$	& $0.916\left(^{+65}_{-70}\right)$		\\
$h_{A_1}(1)$ 		 & $0.906\left(13\right)$		& $0.906\left(13\right)$		\\\hline\hline
%%%
\end{tabular}
}
\caption{Fit results using the BGL (a) and CLN (b) parameterizations. In the BGL fits $a_0^{\mathcal{F}_1}$ is fixed by the value of~$a_0^f$, see Eq.~(\ref{aaa}).\label{tab:fit-results-formfactors}}
\end{center} 
\end{table*}
The $B_c$ resonances also enter the $1^-$ unitarity bounds (see below) as single particle contributions.
The outer functions $\phi_i$ for $i=g, f,\mathcal{F}_1$, can be read from Eq.\,(4.23) in Ref.~\cite{Boyd:1997kz}:
\bea 
\phi_g(z)&=&\sqrt{\frac{n_I}{3\pi \tilde\chi^T_{1^-}(0)}} \frac{2^4 r^2 (1+z)^2 (1-z)^{-\frac12}}{[(1+r)(1-z) +2\sqrt{r}(1+z)]^{4}},\nn\\
\phi_f(z)&=&\frac{4r}{m_B^2}\sqrt{\frac{n_I}{3\pi \chi^T_{1^+}(0)}} \frac{ (1+z) (1-z)^{\frac32}}{
[(1+r)(1-z) +2\sqrt{r}(1+z)]^{4}},\nn\\
\phi_{{\cal F}_1}(z)&=&\frac{4r}{m_B^3}\sqrt{\frac{n_I}{6\pi \chi^T_{1^+}(0)}} \frac{ (1+z) (1-z)^{\frac52}}{
[(1+r)(1-z) +2\sqrt{r}(1+z)]^{5}},\nn
\eea
where $\chi_{1^+}^T(0)$ and $\tilde\chi^T_{1^-}(0)$ are constants given in Table~II, and $n_I=2.6$ represents the number of spectator quarks (three), decreased by a large and conservative SU(3) breaking factor.
Notice that at zero recoil ($w=1$ or $z=0$) there is a relation between two of the form factors
\be
{\cal F}_1(0)= (m_B-m_{D^*}) f(0).\label{aaa}
\ee
The coefficients of the expansions (\ref{BGLexp}) are subject to unitarity bounds based on analyticity and the Operator Product Expansion applied to correlators of two hadronic $\bar c b$ currents. They read \cite{Boyd:1997kz}
\be
\sum_{i=0}^\infty  (a_n^g)^2<1,\quad \sum_{i=0}^\infty \left[(a_n^f)^2+(a_n^{{\cal F}_1})^2\right]<1,
\label {unitcon}
\ee
and ensure a rapid convergence of the $z$-expansion over the whole physical region, $0<z<0.056$. Of course, the series (\ref{BGLexp}) need to be truncated at some power $N$.
In general we find that a truncation at $N=2$ is sufficient for the $|V_{cb}|$ determination, but we have systematically checked the effect of higher orders by repeating the analysis with $N=3,4$, finding very stable results.

The unitarity constraints (\ref{unitcon}) can be made stronger by adding other hadronic channels with the same quantum numbers. For 
instance, the form factor $f_+$ entering the decay $B\to D \ell\nu$ contributes to the left hand 
side of the first equation in (\ref{unitcon}). Since lattice calculations and experimental data 
determine $f_+(z)$ rather precisely~\cite{Bigi:2016mdz}, one can readily verify that 
its contribution is negligible.
More generally, it is well-known that Heavy Quark Symmetry relates the various $B^{(*)}\to D^{(*)}$ form factors in a stringent way: in the heavy quark limit they are all either proportional to the Isgur-Wise function or vanish. These relations can be used to make the unitarity bounds stronger \cite{Boyd:1997kz,Caprini:1997mu}, and to decrease the number of relevant parameters.
The CLN parameterization is built out of these relations, improved with perturbative and $O(1/m) $ leading  Heavy Quark Effective Theory (HQET) power corrections from QCD sum rules, and of the ensuing {\it strong} unitarity bounds.
With respect to the original paper \cite{Caprini:1997mu}, 
the experimental analyses have an additional element of flexibility, as they fit the 
zero recoil value of $R_{1,2}$ directly from data, rather than fixing them at their
HQET values $R_1(1)=1.27$, $R_2(1)=0.80$. 
It is quite obvious that the HQET relations employed in Ref.\,\cite{Caprini:1997mu} have a non-negligible uncertainty. We will not  discuss here how this was estimated and included in \cite{Caprini:1997mu},
but it should be recalled that the accuracy of  the parameterization  for $h_{A_1}(w)$  in Eq.~(\ref{CLN}) was estimated there to be {\it better than 2\%}. Such an uncertainty, completely negligible at the time,
is now quite relevant as can be seen in Eqs.\,(\ref{excl},\ref{incl}). However it  has never been included in the experimental analyses. Similarly, the slope and curvature of 
$R_{1,2}$ in Eq.~(\ref{CLN}) originate from a calculation which is subject to $O(\Lambda^2/m_c^2)$ and $O(\alpha_s \Lambda/m_c)$ corrections and to 
uncertainties in the QCD sum rules on which it is based\footnote{These points are also emphasized in \cite{Bern:2017}, which appeared as we were about to publish this paper on the ArXiv.}.   
 
 The CLN and BGL parameterizations
are both constructed to satisfy the unitarity bounds. They differ mostly in the CLN reliance on next-to-leading order HQET relations between the form factors.  In the following we are going  to verify how important this underlying assumption is for the extraction of $V_{cb}$, remaining mainly agnostic on the validity of the HQET relations, a matter which ultimately will be decided by lattice QCD calculations\footnote{As noted in \cite{Bigi:2016mdz},
recent lattice calculations differ from the HQET ratios of form factors 
at the level of 10\%.}.
Our strategy in the following will be to perform minimum $\chi^2$ fits to the experimental data 
using the CLN or BGL parameterizations; in the latter case we will look for $\chi^2$ minima which respect the constraints (\ref{unitcon}) and evaluate $1\sigma$ uncertainties looking for $\Delta\chi^2=1$ deviations. 
\begin{table*}[t]
\begin{center}
\begin{tabular}{c|c|c}
\hline
\hline
Additional fits 						 	 & $\chi^2/\mathrm{dof}$ 	& $\vert V_{cb}\vert$ 	\\\hline 
%%%
CLN without angular bins    						 	 & $7.1/6$		& $0.0409\left(^{+16}_{-17}\right) $	\\
%%%
BGL ($N=2$) without angular bins 					 	 & $5.1/2$		& $0.0428\left(^{+21}_{-22}\right) $	\\
%%%
CLN only angular bins    					 	 & $23.0/26$		& $ 0.074\left(^{+4}_{-37}\right) $	\\
%%%
BGL ($N=2$) only angular bins   				 	 & $22.3/32$		& $0.058\left(^{+25}_{-31}\right) $	\\
%%%
CLN with $R_{1,2}$ slopes let free 			 	 & $28.1/34$		& $0.0415\left(19\right) $		\\
%%%
%%%
BGL ($N=2$) fit with $R_{1,2}(w=1.4) = $ HQET $\pm 20\%$ (CLN Eq.~(36)) & $31.7/34$		& $0.0407\left(^{+17}_{-18}\right) $	\\\hline\hline
%%%
\end{tabular}
\caption{Additional fits. The  lattice input (11) is always included and LCSR constraints are never included. 
\label{tab:fit-results-formfactors-additional} } 
\end{center}
\end{table*}

\section{Fits and Results}
In our $\chi^2$ fits we use the unfolded differential decay rates measured in Ref.~\cite{Abdesselam:2017kjf}. The Belle Collaboration provides the $w$, $\cos\theta_v$, $\cos\theta_l$, and $\chi$ distributions, measured in 10 bins each, for a total of 40 observables, and the relative covariance matrix. In addition, like Ref.~\cite{Abdesselam:2017kjf}, in the following we always use the value of the form factor $h_{A_1}$  calculated at zero-recoil on  the lattice  \cite{Bailey:2014tva}, 
\begin{align}
h_{A_1}(1) = 0.906 \pm 0.013\,. \label{lattice}
\end{align}
This is the only form factor relevant at zero-recoil, and to the best of our knowledge this Fermilab/MILC calculation  is the only published unquenched calculation. Among older quenched calculations, Ref.~\cite{deDivitiis:2008df} extends up  to $w=1.1$, but  we will not employ it here because of the uncontrolled quenching uncertainty.

\begin{figure*}[ht]
\begin{center}
\subfigure[\label{dGammadwPlot}]{
        \includegraphics[width=0.6\textwidth]{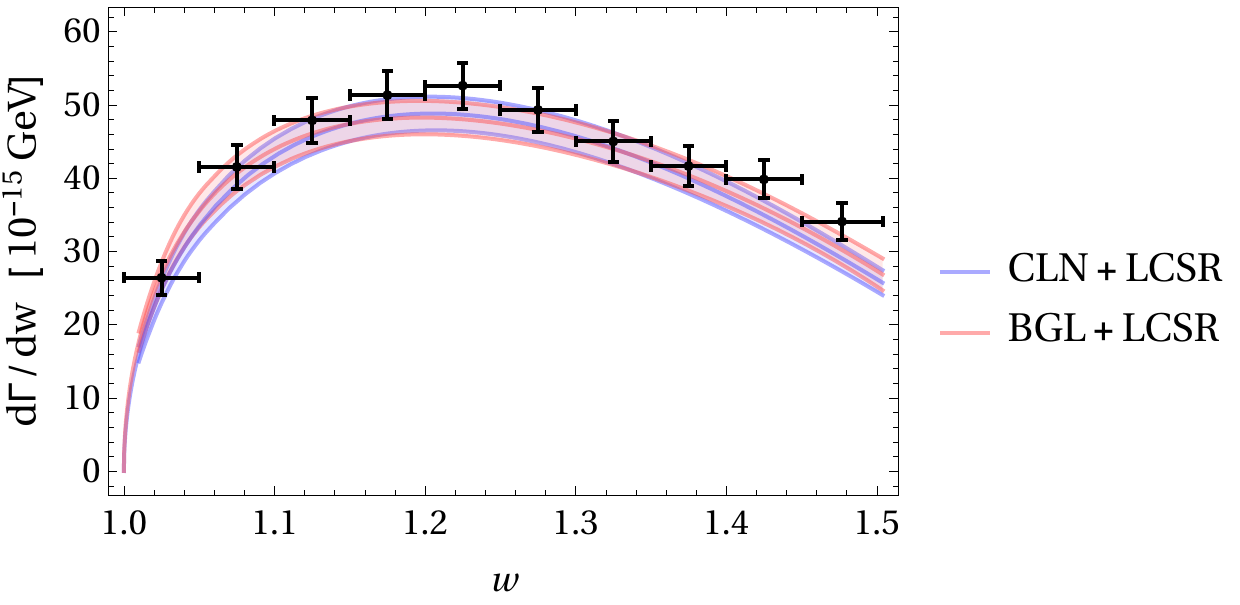}
}
\hfill %
\subfigure[\label{VcbFsqPlot}]{
        \includegraphics[width=0.6\textwidth]{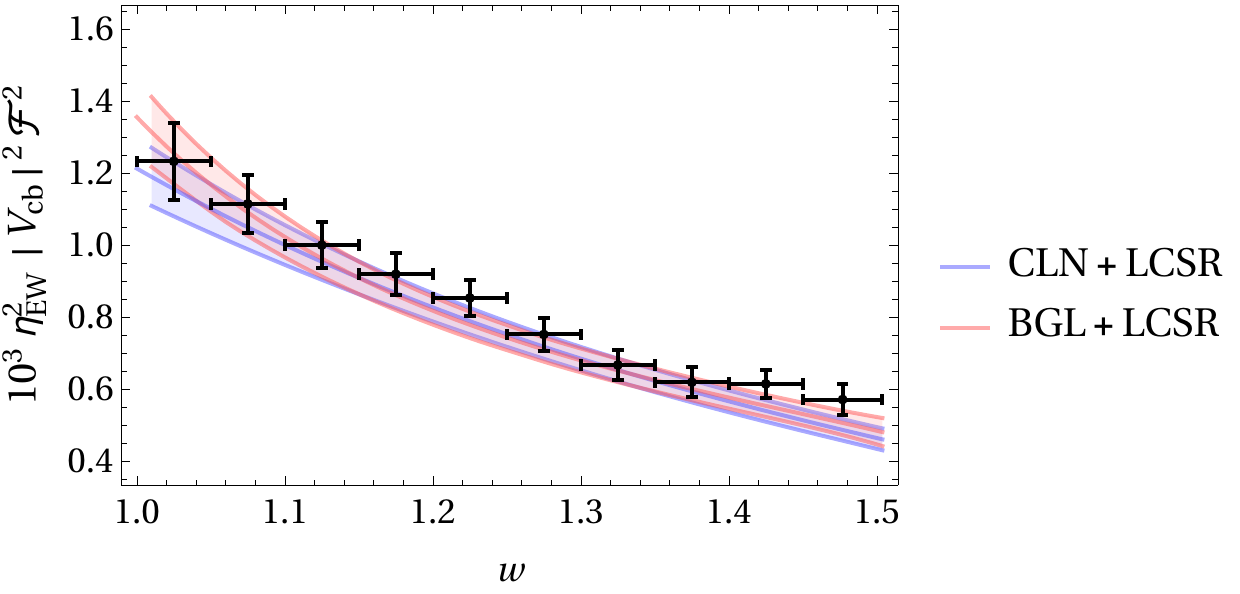}
}
\hfill %
\subfigure[\label{R1R2Plot}]{
        \includegraphics[width=0.6\textwidth]{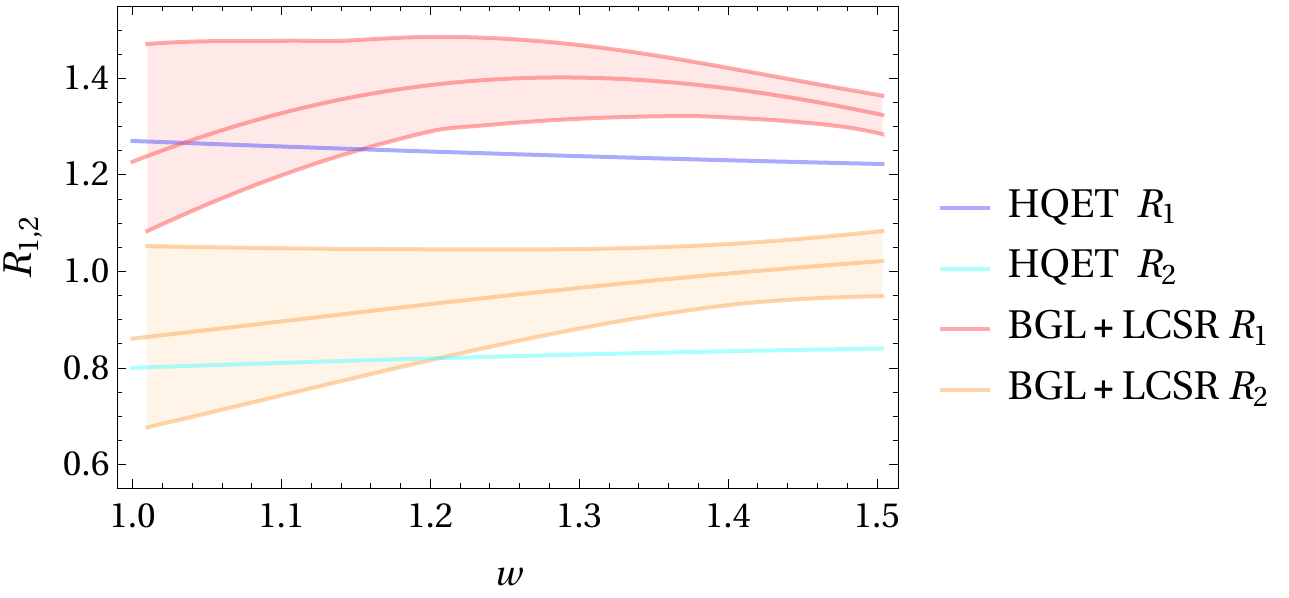}
}
\end{center}
\caption{
Comparison of fit results with different parametrizations. 
\label{fig:fits}
}
\end{figure*}

As far as the determination of $|V_{cb}|$ is concerned, the purpose of a fit to $B\to D^* \ell\nu$ observables is therefore simply to {\it extrapolate} the measurements to the zero-recoil point, where (\ref{lattice}) provides the normalization. As
the differential width vanishes like $\sqrt{w-1}$ as $w\to 1$, see Eq.~(\ref{eq:fully-diff-decay-rate}) and Fig.~\ref{dGammadwPlot}, the extrapolation is not trivial.  Like in the case of $B\to D \ell\nu$, the situation is set to improve significantly as soon as lattice calculations of the form factors at non-zero recoil will become available, but for the moment it is important to keep in mind that the extrapolation should be controlled by the low recoil behaviour of the form factors. 
In this context the angular observables provide very little information, as they are 
integrated over the full $w$ range and receive negligible contribution from the suppressed low-recoil region. Of course, the angular observables are very important to constrain new physics, see for instance Ref.~\cite{Becirevic:2016hea}, but their contribution in the 
determination of $|V_{cb}|$ is marginal.

The results of our BGL and CLN fits to the full data set and to (\ref{lattice}) are given in the first columns of Tables~\ref{tab:fit-results-formfactors}\subref{tab:fit-results-formfactors-BGL} and \ref{tab:fit-results-formfactors}\subref{tab:fit-results-formfactors-CLN}. The results of the CLN fit are in perfect agreement with the 
one in Appendix B of Ref.~\cite{Abdesselam:2017kjf}. Incidentally, Belle's paper also reports the results of a fit performed without unfolding the distributions which gives  $|V_{cb}|=0.0374(13)$. The BGL fit in Table~\ref{tab:fit-results-formfactors}\subref{tab:fit-results-formfactors-BGL}, left column, has a  9\% higher central value and a 40\% larger uncertainty than the CLN fit.
The fits are both good, and such a large shift in $|V_{cb}|$ comes quite unexpected.  We believe it is related to the fact that the CLN parameterization has limited flexibility and that the angular observables dilute the sensitivity to the low recoil region, which is crucial for a correct extrapolation (they also decrease the overall normalisation of the rate by 0.8\%).  This is clearly seen in Fig.1b, where the bands corresponding to the BGL and CLN fits are compared with the data, and one can notice that the CLN band underestimates all the three low recoil points. 
Table \ref{tab:fit-results-formfactors-additional} shows  $|V_{cb}|$  obtained from fits to the $w$ distribution and (\ref{lattice}) only: the CLN fit is 7\% higher and the two parameterizations give consistent results.
Another fit which supports the simple explanation above is one where we give more flexibility to the CLN parameterization, by floating the slopes of the $R_{1,2}$ ratios. The result, shown in Table \ref{tab:fit-results-formfactors-additional},
is again very close to the BGL one.
 \begin{table*}[t]
\begin{center}
\begin{tabular}{c|c|c}
\hline
\hline
Future lattice fits 						 	& $\chi^2/\mathrm{dof}$ 	& $\vert V_{cb}\vert$ 	\\\hline 
%%%
CLN     						 		& $56.4/37$		& $0.0407\left(12\right) $		\\
%%%
CLN+LCSR 					 			& $59.3/40$		& $0.0406\left(12\right) $		\\
%%%
BGL	    					 			& $28.2/33$		& $0.0409\left(15\right) $		\\
%%%
BGL+LCSR   				 				& $31.4/36$		& $0.0404\left(13\right) $		\\\hline\hline
%%%
\end{tabular}
\caption{Fits including an hypothetical future lattice calculation giving  
 $ \frac{\partial \mathcal{F}}{\partial w}|_{w=1} = -1.44\pm 0.07$. 
\label{tab:fit-results-future}} 
\end{center}
\end{table*}

Concerning the quality of the fits we show, one should take into account that all BGL fits are constrained fits where (\ref{unitcon}) are employed after truncation at order $N$. The effective number of degrees of freedom is therefore {\it larger} than the naive counting shown in the Tables (the number of degrees of freedom is not well-defined in a constrained fit, as the parameters are not allowed to take any possible value). This is well illustrated by the second fit in Table~\ref{tab:fit-results-formfactors-additional}, whose $\chi^2/{\rm dof}=5.1/2$ may look suspect. However,  the unitarity constraints play an important role here: without them the best fit would have $\chi^2/{\rm dof}=1.2/2$. 

It can be reasonably argued that the HQET input used in devising the CLN parameterization is important theoretical information that one should not neglect. A simple way to do that is to 
include HQET constraints on $R_{1,2}$ at specific values of $w$ with a conservative uncertainty. As an example, we have used the 
HQET values of $R_{1,2}$ at $w=1.4$ with a 20\% uncertainty in the BGL fit and observed a
downward shift in the $V_{cb}$ central value, see Table \ref{tab:fit-results-formfactors-additional}.  Lowering the uncertainty we observe very little effect: for a 10\% uncertainty,    $|V_{cb}|=0.0407(^{+17}_{-20})$. It turns out that the value of $|V_{cb}|$ depends most sensitively on that of $R_1$ at large $w$.

 Alternatively, one can avoid HQET inputs altogether and employ instead
information on the form factors at maximal recoil from Light Cone Sum Rules \cite{Faller:2008tr}: 
\bea
&h_{A_1}(w_{max})=0.65(18),
\label{LCSR}\\ &R_1(w_{max})=1.32(4),\quad
R_2(w_{max})=0.91(17).\nn
\eea
The results of the BGL and CLN fits with the complete  Belle's dataset and Eqs.\,(\ref{lattice},\ref{LCSR}) are given in Tables \ref{tab:fit-results-formfactors}\subref{tab:fit-results-formfactors-BGL} and \ref{tab:fit-results-formfactors}\subref{tab:fit-results-formfactors-CLN}, right column. The CLN fit is unaffected by the LCSR constraints, while the BGL fit gives a smaller  $|V_{cb}|$. Now the two fits are compatible, but the difference between their $|V_{cb}|$ central values is still larger than 5\%. 
It is interesting to compare $R_{1,2}(w)$ derived from the BGL fit (bands in Fig.~\ref{R1R2Plot})
with the HQET predictions  of the same quantities \cite{Caprini:1997mu} (straight lines).
They are perfectly compatible if one assumes a $\sim 10\%$ uncertainty for the latter.

Finally, we show in Table~\ref{tab:fit-results-future} what would happen if a 5\% determination of the slope  of the form factor ${\cal F}(w)$, see Eq.~(\ref{calF}), at $w=1$ were available from the lattice. For the central value we take the central value of the BGL fit with LCSR constraints.
The results demonstrate the importance  of a precise lattice determination of the slope to control the zero-recoil extrapolation. Indeed, 
the parameterization dependence becomes minimal and the LCSR constraints
become much less important. The quality of the CLN fits deteriorates, while  the BGL uncertainty is still somewhat larger.

%%%%%%%%%%

\section{Final remarks}
 We have performed fits to the recent $B\to D^* \ell \nu$ data by Belle \cite{Abdesselam:2017kjf} with the CLN and BGL parameterizations. The BGL results for $|V_{cb}|$ are consistently higher than those obtained with the CLN parameterization.
One cannot avoid  noticing that the central values of all our BGL fits are  perfectly compatible with (\ref{incl}). However, one should be very careful in interpreting our results: we simply observed that
the Belle data  we have employed lead to different $ |V_{cb}|$ when they are  analysed with two parameterizations which differ mainly in their reliance on HQET relations. The data do not
show any preference for a particular parameterization (both give acceptable fits), but in the absence of new information from lattice on the slope and zero-recoil value of the form factors the BGL parameterization offers a more conservative and reliable choice. It is possible,  even likely, that the behaviour we have observed is accidentally related to the new Belle data only, and that  Babar and previous Belle data would lead to a smaller difference between the CLN and BGL fits.
 Still, we believe that a parameterization that does not incorporate HQET relations but satisfies important unitarity bounds, such as BGL in the way we used it above, would provide a more reliable estimate of the current uncertainty on $|V_{cb}|$.
 
While our findings do not provide a clear resolution of the $|V_{cb}|$ puzzle, they strongly
question the reliability of the current $B\to D^*\ell\nu$ averages \cite{Amhis:2016xyh} and
call for a reanalysis of old experimental data before Belle-II comes into action.

\vspace{2mm}

{\bf Acknowledgements.}
 We are grateful to Florian Bernlochner and Christoph Schwanda for useful communications concerning the Belle Collaboration results.
 
%%%%%%%%%%%%%%%%%%%%
\bibliography{draft_v3.bib}

\end{document}